%% file: maindocument.tex
\documentclass[prd,showpacs,amssymb,amsmath,amsfonts,aps,nofootinbib,notitlepage]{revtex4-1}
\usepackage[top=3cm,bottom=3cm,left=3cm,right=3cm]{geometry}

\usepackage{epsfig}
\usepackage{graphicx}

\usepackage{hyperref}
\hypersetup{colorlinks,citecolor=black,filecolor=black,linkcolor=black,urlcolor=black, 
bookmarksnumbered}

\usepackage{graphicx}
\usepackage{dcolumn}
\usepackage{bm}
\usepackage{amsmath,amsfonts,amssymb,amsthm,amscd,bbm,mathrsfs}
\usepackage{listings}
\usepackage{colortbl}
\usepackage{booktabs}

\newcommand{\mreg}{^{\scriptsize{\textregistered}}}
\newcommand{\be}{\begin{equation}}
\newcommand{\ee}{\end{equation}}

\begin{document}

\title{Optimum Synthesis of Mechanism for single- and hybrid-tasks using Differential Evolution}

\author{F. Pe\~nu\~nuri}
\email{francisco.pa@uady.mx}
%\cortext[cor1]{Corresponding author. Tel.: +52\,999\,9300550 ext. 1051,1052}
\author {R. Pe\'on-Escalante}
\email{rpeon@uady.mx}
\author {C. Villanueva}
\email{cesar.villanueva@uady.mx}
\author {D. Pech-Oy}
\email{davidj@hotmail.com}
\affiliation{Facultad de Ingenier\'ia, Universidad Aut\'onoma de Yucat\'an, A.P. 150, Cordemex, M\'erida, Yucat\'an, M\'exico.}

\begin{abstract}
\input{abstract.tex}

\end{abstract}

\maketitle

\input{introduction.tex}
\input{ClassicalDE.tex}
\input{Notation_conventions.tex}
\input{HybridTask.tex}

\input{18ptsPresc.tex}
\input{18ptsNOpresc.tex}
\input{AckermannSectionE.tex}
\input{Conclusiones}
\input{Acknowledgments.tex}
\input{Apendices.tex}

\bibliography{referencias}

\end{document}

%% file: abstract.tex
%\begin{abstract}

The optimal dimensional synthesis for planar mechanisms using differential evolution (DE) is demonstrated. 
Four examples are included: in the first case, the synthesis of a mechanism for hybrid-tasks, considering path generation, function generation, and motion generation, is  carried out.
The second and third  cases pertain to path generation, with and without prescribed timing. Finally, the synthesis of an Ackerman mechanism is reported.
Order defect problem is solved by manipulating individuals instead of penalizing or discretizing the search space for the parameters.
A technique that consists in applying a transformation in order to satisfy the Grashof and crank conditions to generate an initial elitist population is introduced. As a result, the evolutionary algorithm increases its efficiency.

%\end{abstract}

%% file: introduction.tex
\section{Introduction}
Dimensional synthesis of mechanisms comprises the problems of path, function  and motion generation.
There are three types of methods for this purpose: graphical, analytical, and those involving optimization \cite{Hartenberg1964}.

Graphical methods offer a quick solution by sacrificing accuracy, and are rarely used since computers can do the same work faster and better.

Analytical methods are based on algebraic expressions \cite{Hartenberg1964,Freudenstein1954}, displacement matrix \cite{Suh1978}, complex numbers \cite{Sandor1984}, or continuation methods \cite{Morgan1990} resulting in mechanisms whose error will be zero at the precision points. 

The problem of motion generation, in the case of a planar four-bar mechanism, can be designed based on the Burmester curve.
This is one of the first proposed analytical methods for the dimensional synthesis of mechanisms.
In \cite{Khalid2002} an algorithm for the robust computation of the solution of the five-posed Burmester problem is introduced. In \cite{Bourrelle2007}  a Matlab-based graphical user interface to the algorithm of \cite{Khalid2002} is done.

Also, the general equation of the coupler curve of a four-bar linkage has attracted the attention of researchers. For a given set of points on the coupler curve, Blechschmidt and Uicker \cite{Blechschmidt1986} have used the equation of the coupler curve to synthesize a four-bar linkage by determining the coefficients of the curve. 

The main disadvantage of the analytical methods lies in the maximum number of points of accuracy that can be set. The mechanisms are restricted to move exactly in a number of points equal to the number of independent parameters that define them \cite{Mabie1987,Norton2009}. Even though the mechanisms obtained can reach the precision points, they may have other problems, known as design defects, that are not taken into account during the synthesis process, thereby preventing the mechanisms from fulfilling the task for which they were designed \cite{Mabie1987}. 

Optimization methods are based on numerical methods and allow a large number of design points tolerating a loss of accuracy.
These are formulated in terms of nonlinear programming problems. The optimal solution is found by optimizing an objective function within an iterative procedure. The objective function can be  defined as a difference between the generated and the specified movement, known as the structural error \cite{Suh1978}. In general, it can be defined as the design error, i.e.,  the error that arises when we are trying to satisfy a design equation \cite{Cervantes2009} (which could be the Freudenstein equation). An interesting definition for the objective function is presented in \cite{Singh1997} where it is defined as a kind of entropy that is maximized.  The use of optimization methods is inevitable when the number of positions to be covered during the duty cycle exceeds a certain number (in the case of motion generation synthesis the classical analytical approach is limited to five specified points for a four-bar mechanism).

The interest in optimum synthesis of mechanisms is not new.
There have been a large number of studies on this topic using a variety of methods.
For example, some local search methods have been described in references \cite{Suh1966,Nolle1971,Suh1973,Alizade_Sandor1975, Angeles1988,Akhras_Angeles1990,Liu_Angeles1992,Krishnamurty1992}. The main disadvantages of these methods are that the objective function must be differentiable. Also, they are very sensitive to the initial search point.

Within the global search methods some of the techniques that have been used are Simulated Annealing (SA) \cite{Martinez1998},
%Fuzzy Logic (FL) \cite{Krishnamurty1993},
Neural Network \cite{Vasiliu2001,Galan2009},
Genetic Algorithm (GA) \cite{Fang1994,Kunjur1997,Cabrera2002, Starosta2008,Nariman2009,Acharyya2009},
Particle Swarm Optimization Technique (PSO) \cite{Acharyya2009}, and
Differential Evolution (DE) \cite{Shiakolas2002,Shiakolas2005,Cabrera2007, Acharyya2009,Bulatovic2009}.  
There are works that use a combination of two optimization methods such as
SA-Powell's Method \cite{Ullah1997},
GA-FL \cite{Laribi2004},
Tabu-Gradient \cite{Smaili2005},
Ant Colony Optimization-Gradient (AG) \cite{Smaili2007}, and
GA-DE \cite{Lin2010}.

The use of evolutionary algorithms has been of significant  interest in recent years. For instance,
Ullah and Kota solved the path generation problem by presenting an objective function based on Fourier descriptors that evaluates only the shape differences between two curves \cite{Ullah1997}.
This function is first minimized using a simulated annealing followed by Powell's method.
The size, orientation and position of the desired curve are addressed at a later stage by determining analogous points on the desired and candidate curves.
Similarly, Vasiliu and Yannou \cite{Vasiliu2001} synthesized the dimensions of a planar mechanism whose purpose is to generate a trajectory shape by using a neural network.

Laribi \emph{et al.}  presented the combined GA-FL method to solve the problem of path generation in mechanism synthesis \cite{Laribi2004}. The FL-controller monitors the variation of the design variables during
the first run of the GA and modifies the initial bounding intervals to restart a second run of the GA.

Smaili and Diab (2007)  apply AG  to the mechanism synthesis problem  for single- and hybrid-tasks \cite{Smaili2007}.
Shiakolas introduced a technique called the Geometric Centroid of Precision Points for defining initial bounds for the design variables combining with DE \cite{Shiakolas2002}.

Acharyya and Mandal  carry out the path synthesis of a four-bar linkage using three different methods \cite{Acharyya2009}. They found that the DE  with /rand/1/exp method performs better than the two others; one being a binary-coded genetic algorithm  (BGA) with multipoint crossover, and the other a 
PSO with the constriction factor approach.

%make a comparison among  binary-coded genetic algorithm 
%(BGA) with multipoint crossover, PSO with the constriction factor approach, and a DE  with /rand/1/exp 
%strategy to carry out the path synthesis of a four-bar linkage, resulting in a DE which had the best performance \cite{Acharyya2009}.

In \cite{Lin2010} they used a GA-DE hybrid algorithm to make a path synthesis of a four-bar linkage.
A real-valued genetic algorithm, where the crossover operation of GA is replaced by differential vector perturbation, is employed.

The DE method is a simple yet powerful algorithm for global optimization \cite{Brest2006}. It is not difficult to modify the main operators and try for improvements of the method. In the present work, we use DE to find optimum solutions for the dimensional synthesis problem of four mechanisms. The first three correspond to planar four-bar and the last one to a six-bar mechanism.
The paper is organized as follows: In section \ref{seccionmetodoDE} we present the classical DE method, which is used throughout this work; section  \ref{secnotyconv} presents notation and conventions. 
In Section \ref{sectionhybrid} we employ the idea of hybrid-task for the synthesis of mechanisms as was introduced in \cite{Smaili2007}.
The problem of this section presents us with the difficulty
of mixing angles with lengths. This difficulty is addressed by
introducing a factor that, on the one hand, defines consistently the objective function and on the other hand, allows for proper weighing of  the involved  errors. This is important in order to fulfill the task of function and motion generation.
Moreover, this problem is used to show an easy and effective way to handle the order defect problem. 
The proposed method avoids entirely both individual penalization and space search discretization.

%The proposed method differs from the traditional formalism based on penalization or space %search discretization.

Section \ref{section18presc} deals with the prescribed timing path generation for 18 points and 10 design variables.
We introduce a transformation which constructs an elitist population, in the sense of satisfying the Grashof and crank conditions, avoiding a probabilistic or penalization approach.
This problem has been presented by other authors \cite{Kunjur1997,Cabrera2002,Lin2010}. To avoid some controversies related to the values of the objective function that each of them report, we have written a {\sc Fortran 90} program that evaluates the objective function.

The ideas introduced in  Sections \ref{sectionhybrid} and \ref{section18presc} allow us to solve in a
single manner the path generation problem without prescribed timing
for 18 target points and 27 design variables, which is the
problem described in Section \ref{section18NOpresc}. In Section \ref{Ackermansection}  an Ackerman mechanism is optimized. 
Finally, we present our conclusions in Section \ref{sectionconclu}.

%% file: ClassicalDE.tex
\section{Classical DE}\label{seccionmetodoDE}
\begin{figure}
\begin{center}
\includegraphics[scale=.4]{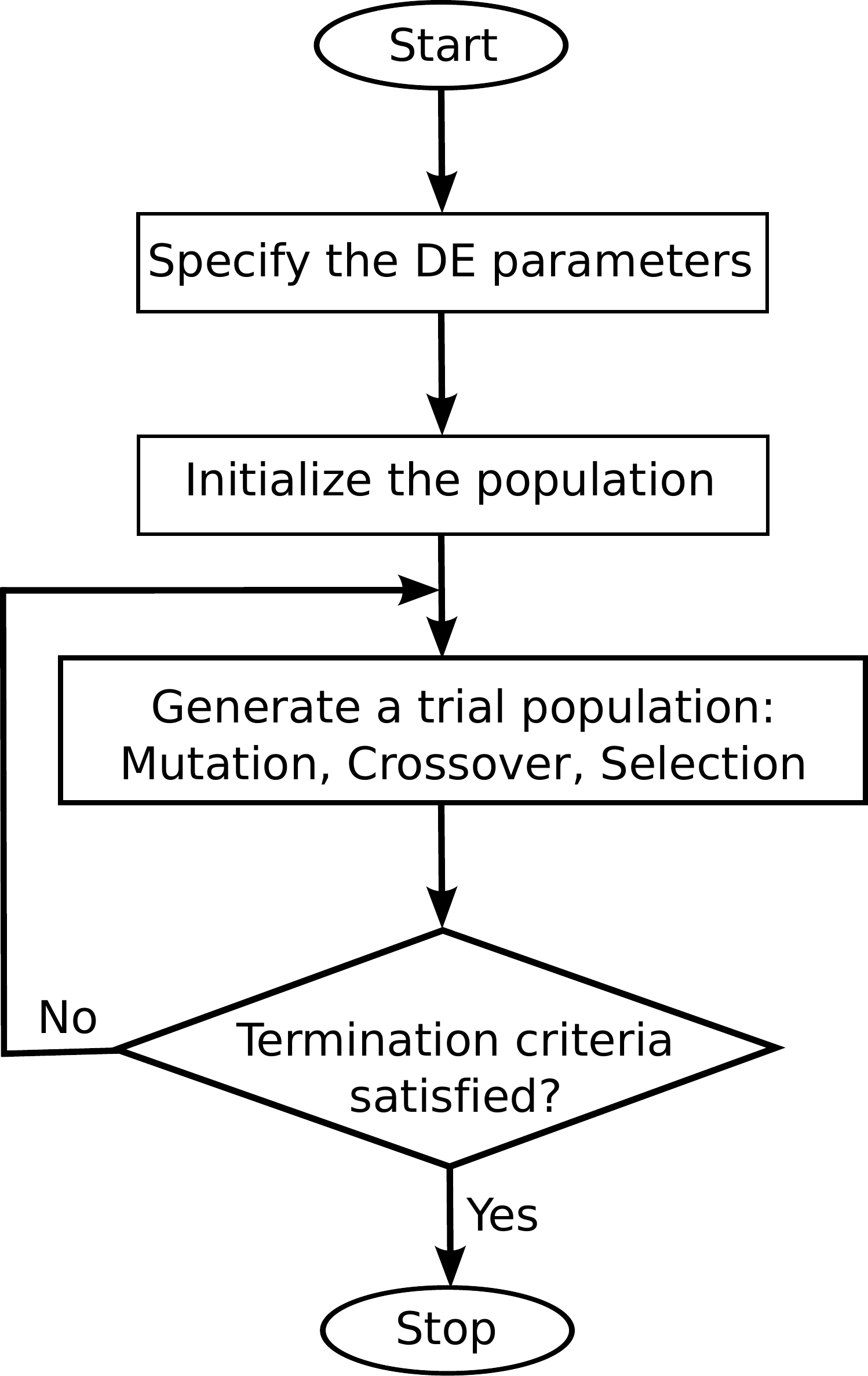} 
\caption{Flowchart for the DE algorithm.}
\label{flowchartDE}
\end{center}
\end{figure}
Below, the original version of the method is outlined \cite{Price_Storn2005}.
\begin{enumerate}
\item 
The population:
\begin{eqnarray}\nonumber \label{ClassicDE}
\mathbf{P_{\mathbf{x},g}} &=&(\mathbf{x}_{i,g}), ~i=1,...m;~~g=0,...g_{\text{max}}\\
\mathbf{x}_{i;g}&=&(x^j_{i;g}), ~j=1,...D;
\end{eqnarray}
where $D$, $m$ and $g_{\text{max}}$ represent the dimensionality of $\mathbf{x}$, the number of individuals and the number of generations respectively.
In \cite{Storn1996} it is mentioned that a good choice for $m$ is $10D$. However, to balance the speed and reliability in \cite{KukonenPrice2005} values from $2D$ to $40D$ are suggested.
\item
Initialization of population:
\begin{equation*}
	x^j_{i;0}=rand^j(0,1)\cdot (b^j_{\text{U}}-b^j_{\text{L}})+b^j_{\text{L}}.
\end{equation*}
Vectors $\mathbf{b}_{\text{U}}$ and $\mathbf{b}_{\text{L}}$ are the parameter limits and
$rand^j(0,1)$ is a random number in $[0,1)$ generated for each parameter.

\item
Mutation:
\begin{equation}
\mathbf{v}_{i;g}=\mathbf{x}_{r_0;g}+F\cdot (\mathbf{x}_{r_1;g}-\mathbf{x}_{r_2;g}).
\end{equation}
The main difference between DE and other evolutionary algorithms like GA comes from this mutation operator. 
$\mathbf{x}_{r_0;g}$ is called the base vector which is perturbed by the difference of other two vectors.

 $r_0, ~r_1,~r_2~\in \{1,2,...m\},~ r_1\neq r_2\neq r_3\neq i$  .
$F$ is a scale factor greater than zero.
Even though upper limits for $F$ do not exist, values greater than 1 are rarely  chosen in the literature \cite{Brest2006,Price_Storn2005,Storn1996,Storn2008}.
\item
Crossover:\\
A dual recombination of vectors is used to generate the trial vector:
\begin{equation}\label{DECrossover}
\mathbf{u}_{i;g}=u^j_{i;g}=\left \lbrace \begin{array}{l} v^j_{i;g}~~\text{if($rand^j(0,1)\leqslant Cr$ or $j=j_{\text{rand}}$)}  \\
                                														x^j_{i;g}~~\text{otherwise.}
 \end{array}\right. 
 \end{equation}
The crossover probability, $Cr \in [0,1]$, is a user-defined value.
\item
Selection:\\
The selection is made according to
\begin{equation}\label{finClassicDE}
\mathbf{x}_{i;g+1}=\left \lbrace \begin{array}{l} \mathbf{u}_{i;g}~~
\text{if $f(\mathbf{u}_{i;g})\leqslant f(\mathbf{x}_{i;g})$}  \\
                                														\mathbf{x}_{i;g}~~\text{otherwise}
 \end{array}\right. 
 \end{equation}
\end{enumerate}

The method just described is known as DE/rand/1/bin. There are
variants of it.  For example, when $F$ is chosen to be a random number,
the variant is called dither. In this work we will use the
exposed method with the dither variant where $F\in  [0; 1)$.  Fig. \ref{flowchartDE} shows the flowchart for the DE algorithm.

%% file: Notation_conventions.tex
%* * * * * * * * * * * * * * * * * * * * * * * * * * * * * * * * * * * *
\section{Mechanism synthesis problem: Notation and conventions}\label{secnotyconv}
%* * * * * * * * * * * * * * * * * * * * * * * * * * * * * * * * * * * *
The simplicity of a 4-bar mechanism, (easy to manufacture and highly reliable) makes it a very important mechanism with a large number of industrial applications. Its use ranges from simple devices such as windshield-wiping mechanisms and door-closing mechanisms to more complicated ones such as rock crushers, sewing machines, round balers, and suspension systems of automobiles \cite{Acharyya2009}.

In this section the notation and conventions used throughout this work are established.
The only exception is in Section \ref{Ackermansection} where we will deal with a 6-bar  mechanism.
%where the used notation and convention are presented within.

\begin{figure}[h!]
	\begin{center}
 		\includegraphics[scale=0.5]{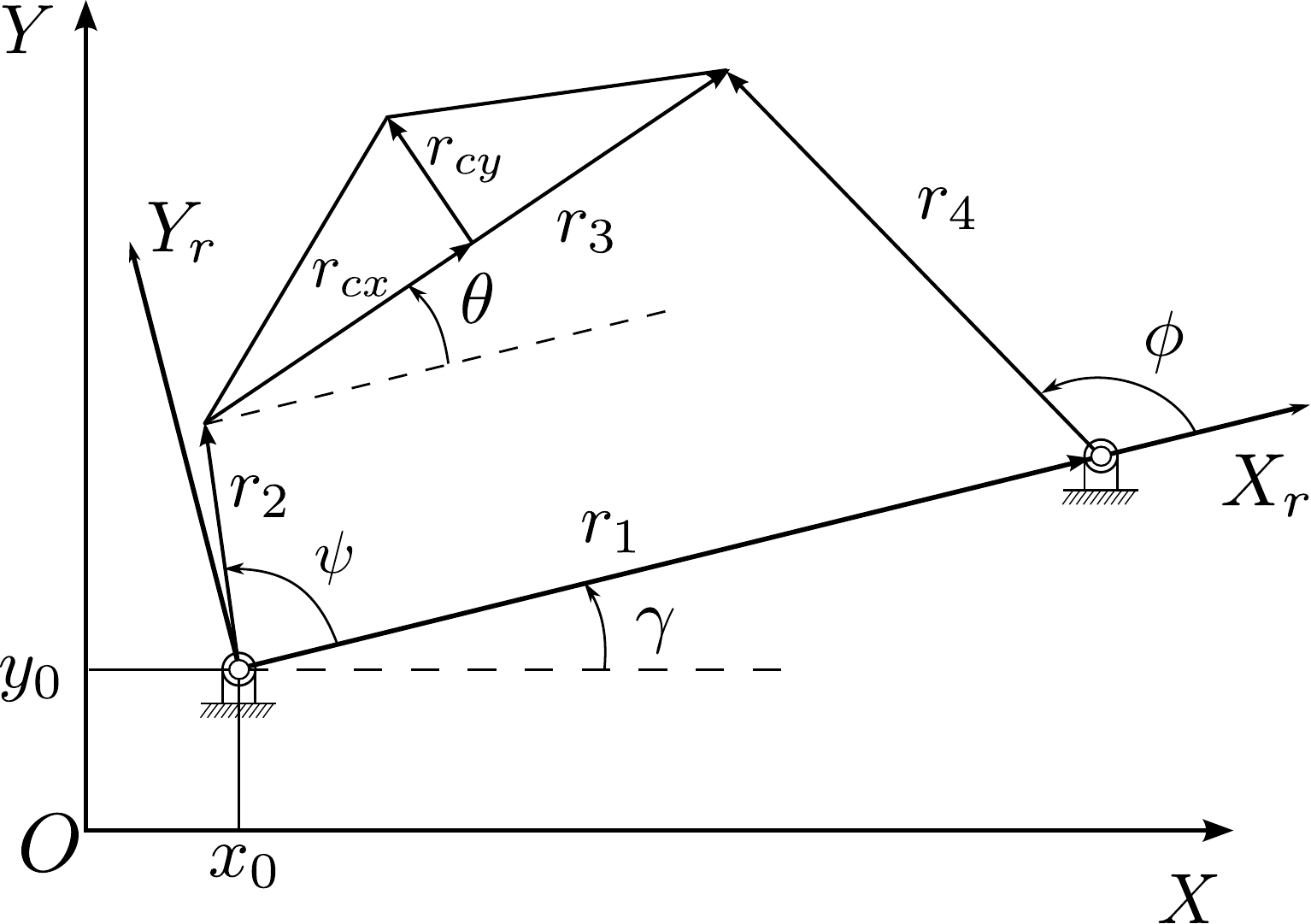}  
 		\caption{Four-bar linkage notation.}
 		\label{wenconv}
 	\end{center}
\end{figure}

A four-bar linkage shown in Fig. \ref{wenconv} consists of four rigid links and four revolute joints.
The set of variables that describes the mechanism (the design variable vector) will be put into the vector $\mathbf{X}$ whose components will be enclosed within braces.
Usually, in the synthesis of a mechanism there are two sets of points (or coordinates), desired and generated points, allocated in the vectors $\mathbf{r}_d$ and $\mathbf{r}_{gen}$, respectively.
A vector error $\mathbf{E}=\mathbf{r}_d-\mathbf{r}_{gen}$ is proposed and the objective function is defined as the square of its Euclidean norm.
\be \label{fobj}
fob=|\mathbf{E}|^2.
\ee
If quantities are not dimensionally homogeneous, constants with appropriate units must be introduced so that equations have compatible units. 
In this work, there are quantities with different units, and some constants are chosen so that $fob$ is dimensionless.
For example, in the problem of motion generation, we have to fit angles and coordinates, so the quadratic  error will be
\begin{equation}\label{mterror}
E^2=\sum_i\left[c(x_{i;d}-x_{i;gen})^2+c(y_{i;d}-y_{i;gen})^2+(\theta_{i;d}-\theta_{i;gen})^2\right],
\end{equation}
where $x_{i;d[gen]}$, $y_{i;d[gen]}$ and $\theta_{i;d[gen]}$ are the coordinates and angles of the desired [generated] point $i$.

%where $x_{i;d}$ ($y_{i;d}$) is the coordinate $x$ $(y)$ of the desired point $i$, $x_{i;gen}$ $(y_{i;gen})$ is the coordinate $x$ $(y)$ of the generated point $i$, $\theta_{i;d}$ is the desired angle $i$ and  $\theta_{i;gen}$ is the generated angle $i$. 
The constant $c$ has a numerical value equal to 1 and is introduced for consistency with units. The objective function is given by
\begin{equation}\label{fobmtdef}
fob=\sum_i\left[f_c^2(x_{i;d}-x_{i;gen})^2+f_c^2(y_{i;d}-y_{i;gen})^2+(\theta_{i;d}-\theta_{i;gen})^2\right].
\end{equation}
The $f_c$ constant is introduced for consistency with units but is not necessarily 1.  Such a constant can be defined by the user as a weight factor. In this work it is chosen as the inverse 
of the longest distance between the coordinates. As a matter of illustration, for the points $P=\{(1,1),(2,3),(-5,-1)\}$ we construct the set
$U_{xy}=\{1,2,3,-5,-1\}$ (i.e., the union of the coordinates) and take $f_c=1/d_r$ with $d_r=\max (U_{xy})- \min (U_{xy})$. In this case  $\min (U_{xy})=-5$, $\max (U_{xy})=3$ thus $f_c=1/8.$ 
%A better choice for the factor $fc$ could be the curvature of the generated trajectory, but in the synthesis of mechanisms this is know  to posteriori. 
Notice that this definition is motivated by the curvature concept. For example, in the case of a circle with radius $r$, we have $s/r=\theta$ or $k s=\theta$ where $k$ is the curvature, $s$ the arc length subtended by the angle $\theta$. 

In general, $ fob \neq 0$ and its minimization process is what generates values for the parameters of a possible mechanism.
In the analysis of mechanisms, two conditions are important. They are known as the crank and Grashof conditions (CG):
%\begin{eqnarray}\label{cmanivela}
%\min(r_1, r_2,r_3,r_4)&=&\text{crank}\\
%\text{crank}  + \max(r_1,r_2,r_3,r_4)& < &\text{sum of the remaining link lengths.} \label{cgrashof}
%\end{eqnarray}
\begin{eqnarray}\label{cmanivela}
\min(r_1, r_2,r_3,r_4)&=&\text{crank},\\
2\,\min(r_1, r_2,r_3,r_4)  + 2\,\max(r_1,r_2,r_3,r_4)& < & r_1+r_2+r_3+r_4. \label{cgrashof}
\end{eqnarray}

In our case $r_2$ is the crank, see Fig \ref{wenconv}.
Whenever we refer to a transformation acting on a vector, $|x\rangle$ is used instead of $\mathbf{x}$.
Usually such transformations are carried out by subroutines or functions in {\sc Fortran 90} and by functions in C++.
For linear transformations, the matrix representation can be used.
In this work, all the algorithms for the synthesis of mechanisms were implemented in {\sc Fortran 90}.  The compiler used was ifort  and the calculations were made in an intel Core 2 Duo processor with velocity of 2.53 GHz, 4 GB of memory and a bus velocity of 1.07 GHz.

%% file: HybridTask.tex
\section{Hybrid task synthesis}\label{sectionhybrid}
%\begin{figure}[h!]
%\begin{center}
% \includegraphics[scale=1]{smailifig2.pdf} 
% \caption{\footnotesize{Convention used in \cite{Smaili2007}.}}
%\label{smailifig}
% \end{center}
% \end{figure}

In this section we analyze the problem presented by McGarva \cite{McGarva1994}. We address the problem from the viewpoint of hybrid tasks as proposed by Smaili and Diab in \cite{Smaili2007}.
The problem has three tasks: function generation, motion generation and path generation.
Table \ref{smailitable}  (as presented in \cite{Smaili2007}) summarizes  the variables used in this study.

\begin{table}[h]
\centering
\scalebox{.9}{
\input{table1.tex}
}
\caption{Hybrid-tasks problem; $\blacktriangle$: Generated crank angle values, *: Non-prescribed values.  }\label{smailitable}
\end{table}

The design variable vector is defined as
\be 
\mathbf{X}=\{x_0,y_0,r_1,r_2,r_3,r_4,r_{cx},r_{cy},\psi_1,\psi_2,\psi_3,\psi_4,\psi_5,\psi_6,\psi_7,\gamma\}.
\ee

In addition to the CG restrictions, we have the following constraints for motion generation:
\begin{eqnarray} \nonumber \label{restricang}
&&\psi_{min}<\psi^j<\psi_{max};~j=\{1,2,...,7\}\\
&&\psi^{k}<\psi^{k+1};~k=\{1,2,..,6\}.
\end{eqnarray}

In this case the objective function consists of three parts:
\begin{equation}
fob=fob_{func}+\widetilde{fob}_{mot}+\widetilde{fob}_{path}
\end{equation}
where, in the usual approach of the least square method, the partial objective functions are defined as:
\begin{align} 
\label{hybmoveq1}
fob_{func}& = \sum_i(\theta_{i;d}-\theta_{i;gen})^2_{func}\\
\label{hybmoveq2}
\widetilde{fob}_{mot} & = \sum_i\left[f_c^2(x_{i;d}-x_{i;gen})^2_{mot}+f_c^2(y_{i;d}-y_{i;gen})^2_{mot}+(\theta_{i;d}-\theta_{i;gen})^2_{mot}\right] \\
\label{hybmoveq3}
\widetilde{fob}_{path} & = \sum_{i}\left[f_c^2( x_{i;d}-x_{i;gen})^2_{path}+f_c^2(y_{i;d}-y_{i;gen})^2_{path}\right]
\end{align}
%In the above equations the generic term  $(\Gamma_{i;d}-\Gamma_{i;gen})^2_{\Upsilon}$ is
%the square of the difference between the desired and the generated points for the part of
%$\Upsilon$ generation. Where, $\Upsilon=func$ stands for function generation, 
%$\Upsilon=mot$ is for motion generation, and $\Upsilon=path$ is for path generation.

The evaluation of the weight factor $f_c$  is explained in Sec. \ref{secnotyconv}. In this case we have $max(U_{xy})=7.03$, $min(U_{xy})=1.22$ thus $f_c^2=0.02$ with units of 
$\text{length}^{-2}$.

The following values were tested for $Cr$: 0.05, 0.1, ..., 0.9. It turns out that 0.3 
gives the best results.
The number of individuals and generations were $m=250$, $g_{\max}=15\,000$, respectively.
The evaluation of $fob$ resulted in a value of $6.99\times10^{-3}$ with the design variables shown in Table \ref{tablahybrid}.

\begin{table}[htb]
\centering
\scalebox{1}{
\input{table2.tex}
}
\caption{Parameter values of an optimal mechanism.  Hybrid-tasks synthesis.}
\label{tablahybrid}
\end{table}

For the searching space we have used the limits:
\begin{align*}
x_{\min}&=\{-15,-15,0,0,0,0,0,0,3,3,3,3,3,3,3,0\}\\
x_{\max}&=\{15,15,15,15,15,15,15,15,5.03,5.03,5.03,5.03,5.03,5.03,5.03,2\pi\}.
\end{align*}
%were found. Secondly, with the obtained results, the searching space was restricted to 
%\begin{align*}
%vx_{\min}&= \{-10,0, 5, 1, 5, 4,10,2,3,3.26,4.36,4.4,4.5,4.5,4.5,5.5\}\\
%vx_{\max}&=\{-7,2,14,2.5,10,10,14,4,3.27,4.3,4.4 ,4.5, 5.03, 5.03, 5.03,6\}.
%\end{align*}

%+ + + + + + + + + + + + + + + + + + + + + + + + + + + + + + + + +
\subsection{Constraints Management}\label{restric}
%+ + + + + + + + + + + + + + + + + + + + + + + + + + + + + + + + +
In this case the CG conditions do not play any active role. We can just verify that they are met after the minimum of $fob$ is obtained. Concerning the requirement of the constraints of Eq. \eqref{restricang}, previous methods are based on the discretization of the search space for $\psi$ angles \cite{Smaili2007},
%Constraints CG are not a problem here. The $fob$ minimum is obtained when they are fulfilled and there is no need to require them because they are verified later.
%Constraints \eqref{restricang} deserve consideration.
%In \cite{Smaili2007} this problem is addressed by discretizing the searching space for $\psi$ angles.
%They use the following discretization approach:
\begin{equation}
\psi^j \in \left[ \psi^j_{\min}, \psi^j_{\max}\right] \, .
\end{equation}
The best fitting angle is selected from this range.
In our case this discretization is not applied, and individuals $\mathbf{x}_{i, g}$ are chosen so as to comply with the restriction of Eq. (\ref{restricang}).
To this end, a random vector of angles within desired limits is generated, and its coordinates written in ascending order. This idea has been used in \cite{Acharyya2009,Lin2010}.
Thus the method here is not exactly a classic DE because the evolution of individuals is manipulated. However, it is clear that the results will be the same. The only thing it does is to accelerate the evolutionary process.
Symbolically, if $|\psi\rangle$ represents a vector of random numbers and $\widehat{sort}$ represents a transformation that puts them in ascending order, then
 \be \label{sorT}
 \psi^j =\left[\widehat{sort}|\psi\rangle\right]^j \, .
 \ee
There are several ways to implement Eq. (\ref{sorT}). In particular, it can be done in the crossover part.
 %Eq. (\ref{DECrossover})
\begin{equation}\label{sortcross}
\mathbf{u}_{i;g}=u^j_{i;g}=\left \lbrace \begin{array}{l}\tilde{ v}^j_{i;g}~~\text{if($rand^j(0,1)\leqslant Cr$ or $j=j_{\text{rand}}$)}  \\
 \tilde{x}^j_{i;g}~~\text{otherwise.}
 \end{array}\right. 
 \end{equation}
where
\be
\tilde{ r}^j=\left[\widehat{sort}|r\rangle\right]^j.
\ee
The transformation $\widehat{sort}$ will act only on those components that we choose to order.
\begin{figure}[h!]
\begin{center}
 \includegraphics[scale=1]{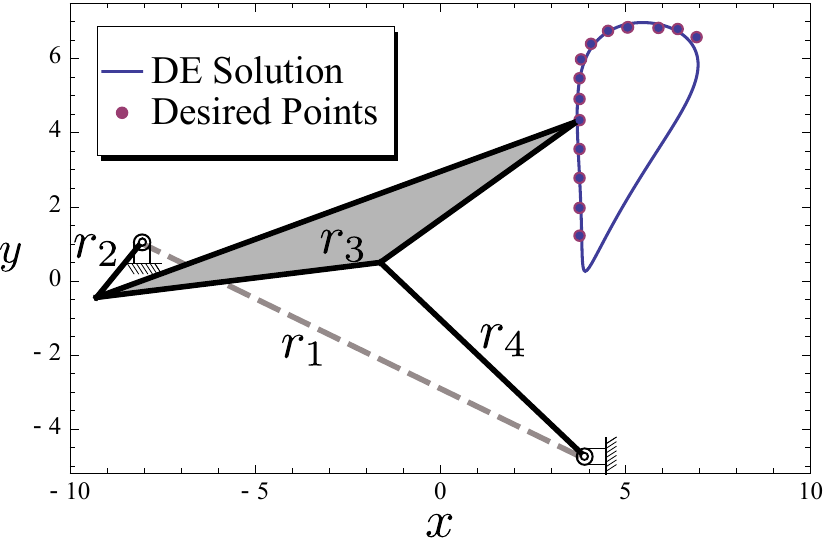} 
 \caption{Optimal mechanism and the corresponding coupler curve. Hybrid-tasks synthesis.}
 \label{hybridtskfig}
 \end{center}
 \end{figure}

For the ordering of the $\psi$ angles, we have used the \emph{heap sort} method \cite{Press1995,Press2002,Press2007} as it is efficient enough and easy to implement.
%Since the number of angles $\psi$ is small, an efficient method for ordering is not required, even so \emph{heap sort} is used \cite{Press1995,Press2002,Press2007}. It is simple to implement and highly efficient.

Penalizing angles $\psi^j$ is not very efficient because the probability of having a set of size $n$ randomly ordered is low if $n$ is large.
For example, the probability to throw in seven random numbers between 0 and 1 (or any other continuum interval) in an ordered way is $1/7!$, which is about $2\times 10^{-4}$. We thus end up with a method without individuals to evolve unless the number of individuals in the initial population were extremely large, which would lead to a grossly inefficient method.
Proceeding as \cite{Smaili2007} is a brilliant possibility and the results so obtained are very good. 
However, discretizing the searching space could prevent us from locating the minimum.
Fig. \ref{hybridtskfig} shows the mechanism obtained.
In \ref{hybanimacion} a program in $Mathematica\mreg$ that makes an animation of the mechanism is shown step by step.

%% file: table1.tex
\begin{tabular}{l c l l l l l l l l l l}
\hline\hline  \\[-1.5ex]
&Desired point, $i$&1&2&3&4&5&6&7&8&9&10\\[0.5ex]
\hline \\[-1.5ex]
Function points&$x_{id}$&7.03& 6.95& 6.77& 6.4& 5.91& 5.43& 4.93& 4.67& 4.38& 4.04\\
&$y_{id}$&5.99& 5.45& 5.03& 4.6& 4.03& 3.56& 2.94& 2.6& 2.2& 1.67\\
&$\psi_i+\gamma$&21 &36 &50 &65 &79 &93 &108 &122 &137 &151\\
&$\theta_i+\gamma$&*&*&*&*&*&*&*&*&*&*\\
&$\phi_i+\gamma$&108&110&113&117&121&126&132&138&143&147\\
&&11&12&13&14&15&16&17\\[0.5ex]
%\cmidrule{3-12} 
\cline{3-12} \\[-1.5ex]
Motion function&$x_{id}$&3.76&3.76&3.76&3.76&3.76&3.76&3.76\\
&$y_{id}$&1.22&1.97&2.78&3.56&4.34&4.91&5.47\\
&$\psi_i+\gamma$&$\blacktriangle$&$\blacktriangle$&$\blacktriangle$&$\blacktriangle$&$\blacktriangle$&$\blacktriangle$&$\blacktriangle$\\
&$\theta_i+\gamma$&$-13$&$-7$&$-2$&2&7&11&14\\
&$\phi_i+\gamma$&*&*&*&*&*&*&*\\
 &&18&19&20&21&22&23&24&25\\[0.5ex]
%\cmidrule{3-12}
\cline{3-12} \\[-1.5ex]
Path point&$x_{id}$&3.8 &4.07 &4.53 &5.07 &5.05 &5.89 &6.41 &6.92\\
&$y_{id}$&5.98 &6.4 &6.75 &6.85 &6.84 &6.83 &6.8 &6.58\\
&$\psi_i+\gamma$&266& 281& 295& 309& 324& 338& 353& 367\\
&$\theta_i+\gamma$&*&*&*&*&*&*&*&*\\
&$\phi_i+\gamma$&*&*&*&*&*&*&*&*\\[0.5ex]
\hline\hline
\end{tabular}

%% file: table2.tex
\begin{tabular}{c c c c c c c c}
\hline\hline  \\[-2.5ex] 
$x_0$&$y_0$&$r_1$&$r_2$&$r_3$&$r_4$&$r_{cx}$&$r_{cy}$\\
 -8.0339& 1.07673& 13.2425& 1.96639& 7.71759& 7.57298& 13.4593& 3.13037\\[0.5ex]
\hline \\[-2.5ex] 
$\psi_1$&$\psi_2$&$\psi_3$&$\psi_4$&$\psi_5$&$\psi_6$&$\psi_7$&$\gamma$\\
 3.5639& 3.83348&4.05641&4.22857& 4.48498&4.71726& 4.92507& 5.83047\\[0.5ex]
\hline\hline
\end{tabular}

%% file: 18ptsPresc.tex
\section{A classical comparison: Path generation for 18 target points and 10 design variables} \label{section18presc}
Recently, in \cite{Lin2010} a hybrid method (GA-DE) was proposed that can synthesize a four-bar mechanism and the problem of prescribed timing path generation for 18 points, 
(previously introduced by \cite{Kunjur1997} and \cite{Cabrera2002}) is addressed. We will optimize this problem by using a DE algorithm.
The objective function value is lower than the reported values of previous references.
It is worth mentioning that the values for the links of the mechanism generated are of the same order of magnitude of the generation path dimensions.

 \begin{figure}[h!]
\begin{center}
 \includegraphics[scale=1]{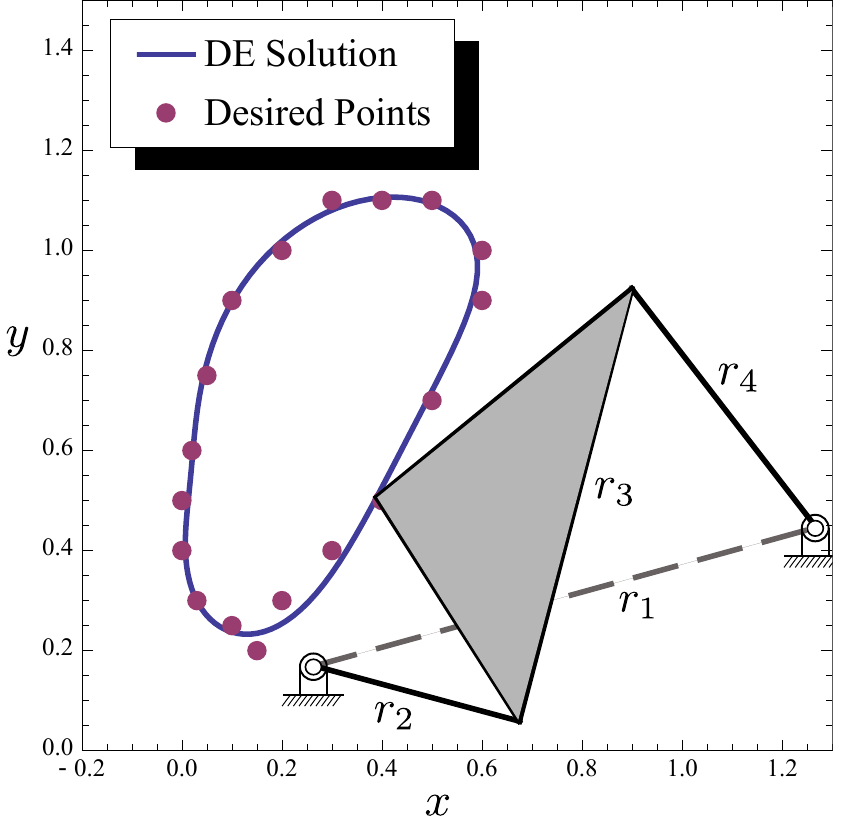} 
 \caption{Optimal mechanism and the corresponding coupler curve. Prescribed timing path generation.}
\label{wenfig}
 \end{center}
 \end{figure}

\subsection{The problem}

The target points are:
\begin{eqnarray}
x_d &=& \{0.5, 0.4, 0.3, 0.2, 0.1, 0.05, 0.02, 0, 0, 0.03, 0.1, 0.15, 0.2,  0.3, 0.4, 0.5, 0.6, 0.6\}\\
y_d &=& \{1.1, 1.1, 1.1, 1, 0.9, 0.75, 0.6, 0.5, 0.4, 0.3, 0.25, 0.2, 0.3, 0.4, 0.5, 0.7, 0.9, 1\}
\end{eqnarray}
The design variable vector is
\be 
\mathbf{X}=\{ x_0,y_0,r_1,r_2,r_3,r_4,r_{cx},r_{cy},\gamma, \psi^0 \}
\ee
and the precribed timing is defined by
\be \label{prescangx}
 \psi^k= \psi^0+ \frac{\pi}{9}(k-1);~k=\{1,2,...,18\}.
\ee
Figure \ref{wenconv} shows the design variables.

This problem has been discussed in Refs. \cite{Kunjur1997,Cabrera2002,Lin2010}.
They do not show the explicit form of $fob$, and there is a controversy concerning the numerical values of the objective function. 
%As an example we show in Table   \ref{tablaerrores} the values  that 
We show in Table   \ref{tablaerrores} the values  that
according to \cite{Lin2010} the other two references should have obtained. 
%The discrepancy is substantial, for instance \cite{Cabrera2002} reports an error of $2.45\times10^{-2}$.

%Wen-Yi Lin \cite{Lin2010} mentioned that there are possible errors in the values reported by \cite{Cabrera2002} (who reports an error of $2.45\times10^{-2}$) and obtains the values shown in Table \ref{tablaerrores}.

\begin{table}[htb]
\centering
\scalebox{1}{
\input{table3.tex}
}\caption{Values for the objective function reported by  \cite{Lin2010}. }\label{tablaerrores}
\end{table}

Here we get the following values of $fob$ for the design variable vectors that they report:
\cite{Lin2010}, $fob=1.0306\times 10^{-2}$; \cite{Kunjur1997}, $fob=1.0214\times 10^{-2}$; \cite{Cabrera2002}, $fob=3.3748\times 10^{-2}$. They are slightly different from the values of \cite{Lin2010}, perhaps because of rounding errors.
With the purpose of avoiding any misunderstanding, in \ref{fobF90} we show a {\sc Fortran 90} program that evaluates $fob$.

Table \ref{tabla18pts1} shows the values for the design variable vector for which the objective function is $9.088\times 10^{-3}$.
The values $0.1$, $0.2$, $\ldots, 0.9$ were tested for $Cr$. It turns out that $0.3$ gives the best results.

%The following values for $Cr$, $0.1, \ldots, 0.9$ were tested, and 0.3 gave the best results.

Figure \ref{wenfig} shows the optimum mechanism and its path.

\begin{table}[htb]
\centering
\scalebox{.95}{
\input{table4.tex}
}
\caption{Parameter  values of an optimal mechanism with $fob=9.088\times 10^{-3}$.}
\label{tabla18pts1}
\end{table}

In order to obtain the last result for $fob$, we proceed in two steps.
First, we choose parameter values inside the interval $[-1.5,1.5]$ for $x_0$ and $y_0$. For the remaining parameters we choose values in $[0,1.5]$, and we evaluate $fob$ over and over until we find a design variable vector for which $fob \leqslant 5\times10^{-2}$.

%First, we look for parameters in $(-1.5,1.5)$ for $x_0$ and $y_0$, and $(0,1.5)$ for the remaining parameters, demanding to stop if $fob \leqslant 5\times10^{-2}$.

Second, from the obtained parameters, the searching space is reduced to 
\begin{align}
vx_{min}& =\{0.2,0.1,0.8,0.3,0.7,0.4,0.2,0.3,0.1,0.7\}\\
 vx_{max}& =\{0.3,0.3,1.1,1.1,1.1,1.1,1.1,1.1,1.1,1.1\}.
\end{align}
The value of $fob = 9.088\times 10^{-3}$ was obtained for 200 individuals and $11\,817$ generations.
We could reach smaller values of $fob$ if the generation number and/or individual number were increased, but improvements are not considerable. For example, for 200 individuals and $30\,000$ generations we obtain $fob=9.06\times 10^{-3}$. Moreover, by making a third refinement of the searching space, we obtain $fob=9.03\times 10^{-3}$ for the design variables shown in Table \ref{tabla18pts2}.

\begin{table}[htb]
\centering
\scalebox{.95}{
\input{table5.tex}
}
\caption{Parameter  values of an optimal mechanism with $fob=9.03\times 10^{-3}$.}
\label{tabla18pts2}
\end{table}

\subsection{On steps 1 and 2}\label{etapas12}
In this work we subdivide the optimization task in two steps. In the first step we use an elitist population in the sense of choosing only those individuals that satisfy the CG condition. To this end we construct a transformation that takes an individual that does not
satisfy the CG condition and turns it into one that does. Then, in the second stage (with the result for the possible mechanism obtained in this first stage) we refine the searching space, remove the CG condition and re-run the 
optimization program. The process terminates when some criteria have been met and the individual satisfies the CG condition. 

\subsection{On the construction of the elitist population}\label{Grshmanip}
In order to construct individuals satisfying the CG conditions we could proceed in a random way, but this would be inefficient. In this work we proceed as follows:

Assuming the links of the four-bar mechanism belong to the interval (0,1), four random numbers are generated (the links) and they are sorted in ascending order. At this point, proceeding randomly would not be a bad choice since the probability of satisfying the CG condition for the sorted list is 0.5.
However a better choice will be to construct a transformation $\hat{T}$ that makes the CG condition fulfill the ascending order list, $|x\rangle$.
There are many possible forms for the transformation $\hat{T}$. We define $\hat{T}=\hat{F}\hat{R}$, where $\hat{R}$ is defined as the transformation that inverts the components of a vector and $\hat{F}$ a reflection plus a translation.
Symbolically,
\begin{align} 
\hat{R}|x_1,x_2,x_3,x_4\rangle & =|x_4,x_3,x_2,x_1\rangle, \\
\hat{F}|x\rangle & =-|x\rangle +|1\rangle, \\
|1 \rangle & =|1,1,1,1\rangle \, .
\end{align}
If the upper limit for the links is $|L\rangle$, we replace $|1\rangle$ by $|L\rangle$.

Notice that $\hat{R}$ is a linear transformation,  whereas $\hat{F}$ is not.
Once the vector that satisfies the Grashof condition has been constructed, the crank is taken as the lesser of the elements; thus the conditions CG will be satisfied.

For example, suppose that we have the four numbers 
${\mathbf x}_r = \{0.38, 0.98, 0.25, 0.19\}$ which do not satisfy the CG condition. After sorting them we have $|x\rangle=|0.19,0.25,0.38,0.98\rangle$  and
\begin{align*} 
 \hat{R}| x \rangle&=|0.98,0.38,0.25,0.19\rangle,\\ 
\hat{F} \hat{R}| x \rangle&=|-0.98,-0.38,-0.25,-0.19\rangle + |1,1,1,1\rangle,\\
\hat{F} \hat{R}| x \rangle&=|0.02,0.62,0.75,0.81\rangle.
\end{align*} 
By choosing the crank as 0.02, we can see that ${\mathbf x}_g=\{0.02,0.62,0.75,0.81\}$ satisfies the CG conditions since $\min(xg)=0.02$, $\max(xg)=0.81$ and $0.02+0.81 < 0.62+0.75.$

In general, suppose we have four positive numbers less or equal than 1 that are  sorted in ascending order, but that do not satisfy the CG conditions. 
Let $|x\rangle=|x_1,x_2,x_3,x_4\rangle$ be the vector containing such numbers.
We have $x_1+x_4>x_2+x_3$ since the numbers are sorted and  do not comply   Eq. (\ref{cgrashof}). Clearly $-x_1-x_4<-x_2-x_3$ and $(1-x_1)+(1-x_4) <(1-x_2)+(1-x_3)$. For these four constructed numbers,  the minimum is $(1-x_4)$  and the maximum is $(1-x_1)$ so the CG conditions are satisfied if we chose the crank as $(1-x_4)$.

%% file: table3.tex
\begin{tabular}{c c c}
\hline\hline  \\[-2.ex]
  \textbf{Wen-Yi Lin \cite{Lin2010} }     & \textbf{Kunjur and Krishnamurty \cite{Kunjur1997}}    & \textbf{Cabrera etal \cite{Cabrera2002}}  \\[0.5ex]
    \hline \\[-2.ex]
    $fob=1.08613\times 10^{-2}$  &  $fob=1.09034 \times 10^{-2}$ & $fob=3.48391\times 10^{-2}$ \\[0.5ex]
\hline\hline
\end{tabular}

%% file: table4.tex
\begin{tabular}{c c c c c c c c c c}
\hline\hline  \\[-2.ex] 
$x_0$&$y_0$&$r_1$&$r_2$&$r_3$&$r_4$&$r_{cx}$&$r_{cy}$&$\gamma$&$\psi^0$\\
 0.27892& 0.11673& 1.08913& 0.42259& 0.96444& 0.58781& 0.39137& 0.42950& 0.32195& 0.86323\\[0.5ex]
\hline\hline
\end{tabular}

%% file: table5.tex
\begin{tabular}{c c c c c c c c c c}
\hline\hline  \\[-2.ex] 
$x_0$&$y_0$&$r_1$&$r_2$&$r_3$&$r_4$&$r_{cx}$&$r_{cy}$&$\gamma$&$\psi^0$\\
 0.26439& 0.16956& 1.04028& 0.42446& 0.89397& 0.60308& 0.36129& 0.38864& 0.26873& 0.90493\\[0.5ex]
\hline\hline  
\end{tabular}

%% file: 18ptsNOpresc.tex
\section{Path generation without prescribed timing for 18 target points and 27 design variables}\label{section18NOpresc}
It is interesting to synthesize  the above mechanism without the prescribed timing Eq. (\ref{prescangx}). Finding a minimum for the objective function is now more difficult. We have 27 design variables and the order defect problem appears hard to solve. 

Let 
\begin{align}\nonumber
\mathbf{X} = &\{ x_0,y_0,r_1,r_2,r_3,r_4,r_{cx},r_{cy},\gamma, \psi^0 ,\psi^1,\psi^2, \psi^3, \psi^4, \psi^5, \psi^6, \psi^7, \psi^8, \psi^9, \psi^{10}, \psi^{11}, \psi^{12}, \psi^{13}, \psi^{14},\\
& \psi^{15}, \psi^{16}, \psi^{17} \}
\end{align}
be the design variable vector. Besides the CG conditions we also have the requirement
\begin{eqnarray} 
&&\psi^{k}<\psi^{k+1};~k=\{0,1,..,16\} \label{orden18psi}
\end{eqnarray}
which prevents the order defect.

The use of penalization for the restriction Eq. (\ref{orden18psi}) is not effective. Practically all the individuals would be penalized as the probability of finding one that would not is 1/18! --\,a very small probability. If we discretize the searching space for angles then  there is no guarantee that the minimum will lie in the generated intervals. However, if we adopt the approach stated in subsection \ref{restric}, the problem is easily solved and in a consistent manner.

The time used for the algorithm was 110 seconds and this was the longest time for all the programs run in this study. The running times for the other cases, were between 30 and 80 seconds. The value of the objective function was $fob=3.69 \times 10^{-3}$ for the design variable vector whose components are shown in Table \ref{section18NOpresc}.

\begin{table}[htb]
\centering
\scalebox{1}{
\input{table6.tex}
}
\caption{Parameter values of an optimal mechanism. Path generation without prescribed timing.}
\label{table18Nopres}
\end{table}

The searching interval for the angles $\psi$ was
\be
0<\psi^j<2\pi;~j=\{0,1,...,17\}.
\ee
It is well known that DE can yield individuals that do not belong to the searching interval. This is the case for the last angle. However, since there is no order defect the values of table \ref{table18Nopres} are an acceptable solution for the problem.

Once again the result is obtained in two steps. First, we choose an elitist initial population that  satisfies the CG conditions. Then, the CG condition is removed and the searching space is restricted according to the solution obtained in the first step.

It is worthwhile to mention that we tried to optimize the $fob$ function using the DE method without the transformations of sections \ref{restric} and \ref{Grshmanip} but the method was not capable of finding the minimum.

%% file: table6.tex
\begin{tabular}{c c c c c c c c c c}
\hline\hline  \\[-2.5ex]  
$x_0$&$y_0$&$r_1$&$r_2$&$r_3$&$r_4$&$r_{cx}$&$r_{cy}$&$\gamma$\\
0.22922& -0.63525& 2.27468& 0.44667& 2.18422& 0.72409& 1.02937&0.82440& 0.58183\\[0.5ex]
\hline  \\[-2.ex] 
$\psi^0$ &$\psi^1$&$\psi^2$& $\psi^3$& $\psi^4$& $\psi^5$& $\psi^6$& $\psi^7$& $\psi^8$\\
0.78140& 1.09985& 1.34998& 1.68045& 2.00009&2.35036& 2.70304& 2.95102& 3.22683\\[0.5ex]
\hline  \\[-2.ex]  
$\psi^9$& $\psi^{10}$& $\psi^{11}$& $\psi^{12}$& $\psi^{13}$& $\psi^{14}$ & $\psi^{15}$& $\psi^{16}$& $\psi^{17}$ \\
3.58801&4.11376& 4.35829&4.70801& 5.07939& 5.35914& 5.76271& 6.21586& 6.49216\\[0.5ex]
\hline\hline
\end{tabular}

%% file: AckermannSectionE.tex
\section{Ackerman steering linkage synthesis}\label{Ackermansection}
In this section DE is used for the synthesis of an Ackerman steering. For the deduction of the equations used and a detailed treatment of the problem see \cite{Jazar2008}.

\begin{figure}[h]
\begin{center}
\includegraphics[scale=0.7]{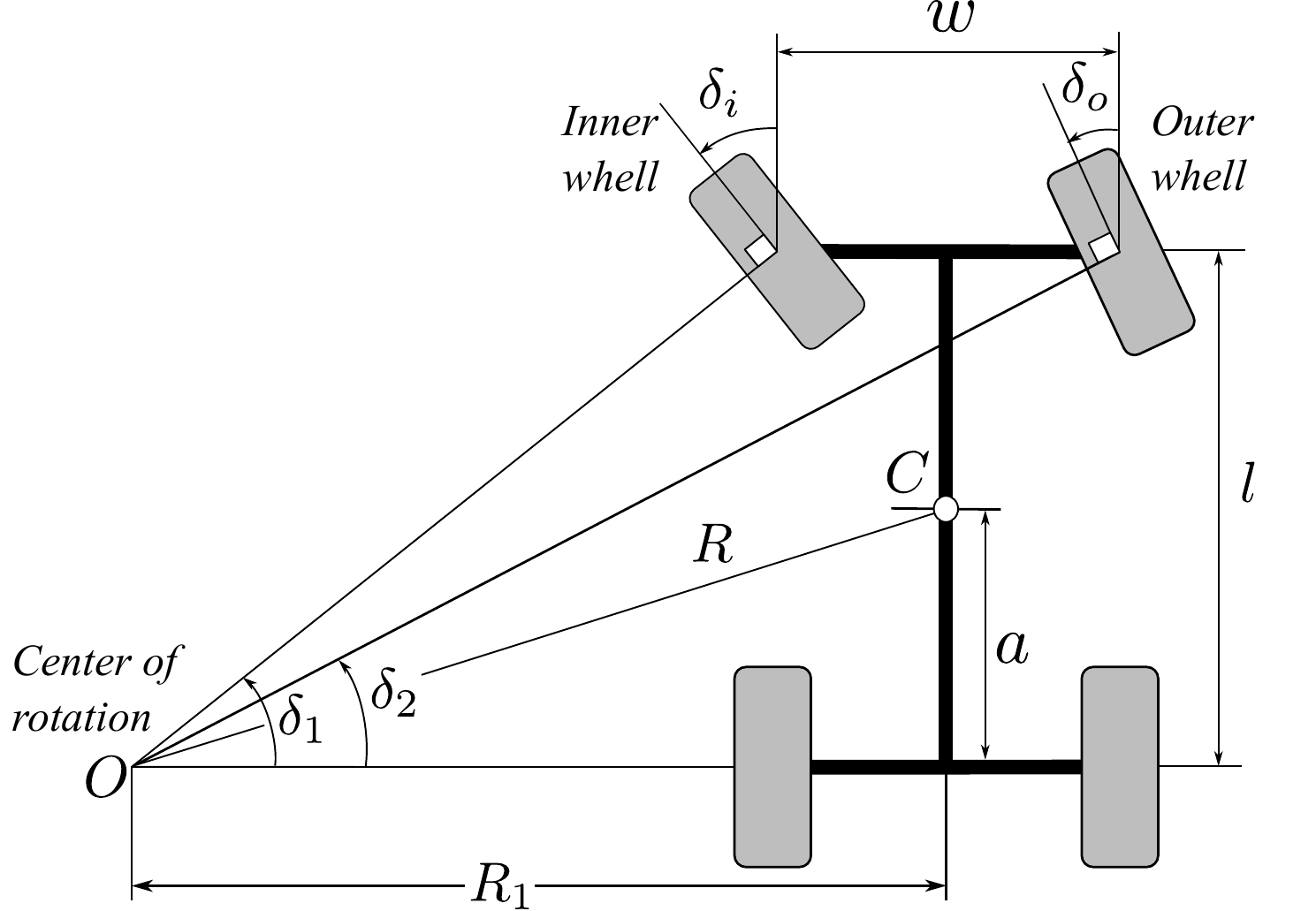}
\caption{Vehicle diagram.}
\label{fig:carro}
\end{center}
\end{figure}

It is known that when a vehicle is moving very slowly there is a kinematic condition between the inner and outer wheel that allows it to turn slip-free. The condition is called the Ackerman condition and is written as follows:
\be
\cot{\delta_o}-\cot{\delta_i}=\frac{w}{l},
\label{ecua:ackerman}
\ee
where $w$ and $l$ represent the width and length of the vehicle, $\delta_o$ and $\delta_i$ are the rotation angles of the wheels (Figure \ref{fig:carro}).

In general it is desirable for a mechanism to satisfy the Ackerman condition. Unfortunately, there is no four-bar mechanism that can fulfill the Ackerman condition perfectly. However, it is possible to synthesize a six-bar mechanism to work closely to the Ackerman condition and be exact at a few points.

\begin{figure}
\begin{center}
\includegraphics[scale=1]{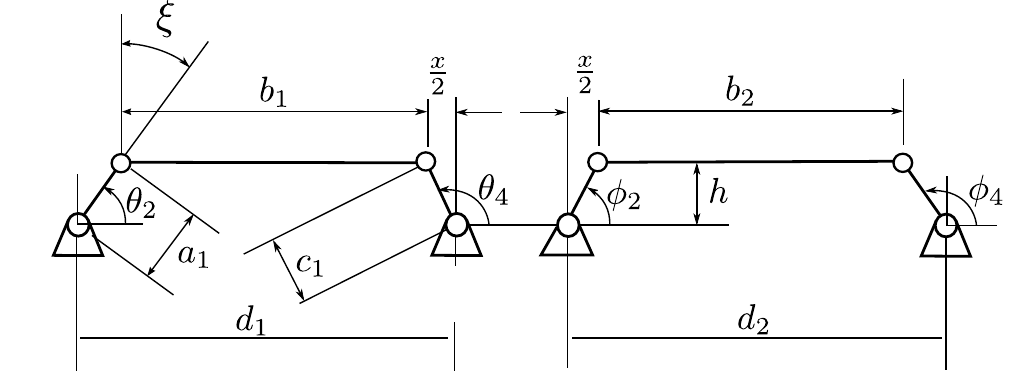}
\caption{Six-bar Watt's mechanism.}
\label{ackerman}
\end{center}
\end{figure}

A six-bar Watt's mechanism can be used to design the vehicle steering. The sizes In this case are $w=1$ m, $l=1.8$ m and the minimum radius $R=2.5$ m. The position of the center of mas with respect to the rear axle is $a=0.45$ m.
 
We have
\be
R=\sqrt{a^2+l^2\cot^2{\delta_M}},
\label{ecua:R}
\ee
with
\be
\delta_M=\frac{\cot \delta_o + \cot \delta_i}{2} 
\ee
therefore $\delta_M=37.2731^\circ$, $R_1=l\cos{\delta_M}$ and consequently $R_1=2.36514$ m. 

From trigonometry we have
\begin{eqnarray}
\delta_i=\arctan\left(\frac{l}{R_1-\frac{w}{2}}\right);~~\delta_o=\arctan\left(\frac{l}{R_1+\frac{w}{2}}\right)
\label{eq:delta}
\end{eqnarray}
so we obtain that  $\delta_i$ and $\delta_0$ must lie in the ranges  $-32.1387^\circ \leqslant \delta_i\leqslant 43.9818^\circ$ and $-43.9818^\circ \leqslant \delta_o \leqslant 32.1387^\circ$ in order to achieve the desired turning radius.

Unlike previous examples where the number of points is finite, in this case it is possible to use an arbitrary number of desired angles. Therefore, it is convenient to define the objective function as
\be
fob=\frac{|\mathbf{E}|^2}{n},
\ee
where $\mathbf{E}$ is the vector containing the $n$ differences between $\delta_2$ and $\delta_{ack}$. The angle $\delta_{ack}$ is the steering angle $\delta_o$ from Eq. (\ref{ecua:ackerman}).

\begin{table}
\centering
\input{table7.tex}
\caption{Parameter values for the multi-link Ackerman steering.}
\label{tab:ackermansteeering}
\end{table}

\begin{figure}
\begin{center}
\includegraphics[scale=1.2]{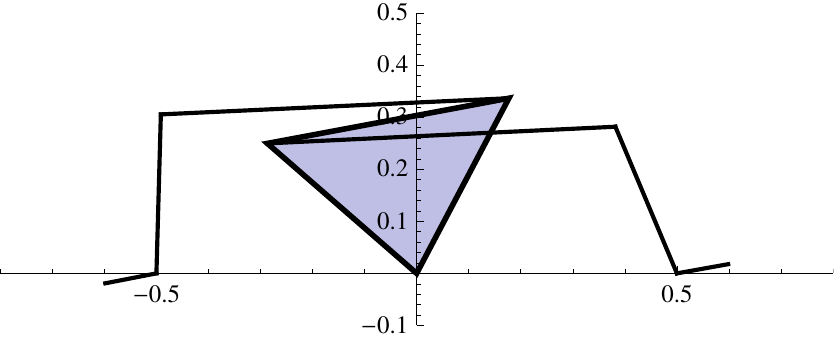}
\caption{ Optimal mechanism. Ackerman steering linkage synthesis.}
\label{obtenido}
\end{center}
\end{figure}

\begin{table}[h]
\centering
\input{table8.tex}
\caption{Desired ($\delta_{ack}$) and generated ($\delta_2$) angles. }
\label{tabladeltack}
\end{table}

The design variable vector is
\be
\mathbf{X}=\{h,x,\xi\} \, .
\ee
Fig. \ref{ackerman} shows the mechanism.

With a search space of $0.1 \leqslant h \leqslant 0.45$, $-0.5 \leqslant x \leqslant 0.2$, $13^\circ\leqslant \xi\leqslant30^\circ$,
and a working range $(-35^\circ,~45^\circ)$ with steps of $0.1^\circ$ for $\delta_1$ we obtain the values shown in Table \ref{tab:ackermansteeering} for the design variables. The objective function is found to be $fob=7.6\times 10^{-5}$. 

The obtained mechanism is illustrated in Fig. \ref{obtenido}. Table \ref{tabladeltack} shows the values of the desired angles and generated angles.

%% file: table7.tex
\begin{tabular}{c c c}
\hline\hline  \\[-2.ex] 
$h$&$x$&$\xi$\\[0.5ex]
\hline \\[-2.ex] 
0.298192&-0.472091&0.219837\\[0.5ex]
\hline\hline
\end{tabular}

%% file: table8.tex
\begin{tabular}{c c c}
\hline\hline  \\[-2.ex]  
$\delta_1$     & $\delta_{ack}$&$\delta_2$\\[0.5ex]
 \hline  \\[-2.ex]
  $-30^\circ$  & $-40.471^\circ$& $-40.254^\circ$\\
  \midrule $-20^\circ$  & $ -24.567^\circ$&$ -23.700^\circ$\\
  \midrule  $-10^\circ$ & $-11.069^\circ$&$ -10.810^\circ$\\
  \midrule $0^\circ$     &  $0^\circ$& $0^\circ$\\
  \midrule $10^\circ$  &   $ 9.117^\circ$& $9.303^\circ$\\
  \midrule $20^\circ$  &   $16.822^\circ$& $17.332^\circ$\\
  \midrule $30^\circ$  &   $23.571^\circ$& $24.174^\circ$\\
  \midrule $40^\circ$  &   $29.720^\circ$& $29.869^\circ$\\[0.5ex]
  \hline\hline
\end{tabular}

%% file: Conclusiones.tex
\section{Conclusions}\label{sectionconclu}

Dimensional synthesis of mechanisms is a subject of great relevance in
the field of mechanical design. Among the great variety of
optimization methods available, those that employ evolutionary
algorithms have seen an increase in use due to the excellent  results that they allow.

In this work we have presented a methodology that uses differential
evolution to solve the dimensional synthesis problem of four mechanisms. With the use of a heuristic deduction, we have determined a weight factor that allows us to solve the hybrid-tasks problem in an
efficient manner. Two transformations were implemented in the differential evolution algorithm. The first one deals with the order defect problem and was coded in the crossover part of the differential evolution algorithm. With this transformation, the penalization approach and the use of big populations are avoided. In addition, the chance of not finding the minimum of the objective function has disappeared as the need of discretization of the search space is also avoided. 
The second transformation constructs elitist populations in the sense that their individuals satisfy the Grashof and crank  conditions. Therefore, a random generation and/or  a penalization procedure are avoided, which makes this method more efficient.

%Such transformation  are quite easy to implement in the differential evolution algorithm but they can also be  adapted to other
%evolutionary algorithms.

%With the above ideas, all the problems that the document present are solved in a good manner but one is worthwhile to mention, the path generation without prescribed timing for the crank angles with 18 target points. It is important to mention that without such transformations the differential evolution algorithm fails in finding the minimum for the objective function.

Something that deserves mention is the amazing speed of convergence of
the differential evolution method which for generations as large as $80\,000$, the total CPU time
was less than two minutes in a single processor.

%% file: Acknowledgments.tex
%\vspace*{1cm}
%{\bf{Acknowledgments}}
%* * * * * * * * * * * * * * * * * * * * * * * * * * * * * * * * * * * *
\section*{Acknowledgments}

We want to thank F. Larios for proofreading the manuscript. We also thank PROMEP and Conacyt for support.

%% file: Apendices.tex
\appendix
%* * * * * * * * * * * * * * * * * * * * * * * * * * * * * * * * * * * *
\section{{\sc Fortran 90}  objective function  ($fob$). Path generation for 18 target points and 10 design variables}\label{fobF90}

What follows is the code for the $fob$ function in {\sc Fortran 90}. 
%We have decided to use the negative branch for the theta angle and let the DE algorithm do  its job.
\begin{verbatim}
Double precision function fob(x0,y0,r1,r2,r3,r4,rcx,rcy,gamma,psi0) 
Implicit None
Integer, Parameter:: Np=18
Double precision, Parameter :: Pi=3.14159265358979d0
Double precision :: x0,y0,r1,r2,r3,r4,rcx,rcy,gamma,psi0,L1,L2,L3,xd(Np), &
yd(Np),psi(Np),KA(Np),KB(Np),KC(Np),theta(Np),px(Np),py(Np),Ex(Np),Ey(Np),&
Ex2,Ey2
Integer :: k

xd=(/0.5d0, 0.4d0, 0.3d0, 0.2d0, 0.1d0, 0.05d0, 0.02d0, 0d0, 0d0,0.03d0,&
0.1d0, 0.15d0, 0.2d0, 0.3d0, 0.4d0, 0.5d0, 0.6d0, 0.6d0/)

yd = (/1.1d0, 1.1d0, 1.1d0, 1d0, 0.9d0, 0.75d0, 0.6d0,0.5d0,0.4d0,0.3d0,&
0.25d0, 0.2d0, 0.3d0, 0.4d0, 0.5d0, 0.7d0, 0.9d0, 1d0/)

L3=(r4**2 - r1**2 - r2**2 - r3**2)/(2d0*r2*r3)
L2=r1/r3
L1=r1/r2

Do, k=1,Np
psi(k) = psi0 + (k-1)*Pi/9d0
Enddo

KA = Dcos(psi) - L1 + L2*Dcos(psi) + L3
KB = -2d0*Dsin(psi)
KC = L1 + (L2 - 1)*Dcos(psi) + L3

theta = 2d0*Datan2(-KB - Dsqrt(KB**2-4d0*KA*KC),2d0*KA) 

px = x0 + Dcos(gamma)*(r2*Dcos(psi) + rcx*Dcos(theta) - rcy*Dsin(theta)) - &
	Dsin(gamma)*(r2*Dsin(psi) + rcx*Dsin(theta) + rcy*Dcos(theta))
	
py = y0 + Dsin(gamma)*(r2*Dcos(psi) + rcx*Dcos(theta) - rcy*Dsin(theta)) + &
Dcos(gamma)*(r2*Dsin(psi) + rcx*Dsin(theta) + rcy*Dcos(theta))

Ex = xd-px
Ey = yd-py

Ex2 = Dot_Product(Ex,Ex)
Ey2 = Dot_Product(Ey,Ey)

fob = Ex2 + Ey2
Return
End
\end{verbatim}

\section{\textit{Mathematica}$\mreg$ hybrid task animation}\label{hybanimacion}
% Notice Figs. \ref{wenconv} and \ref{smailifig} for conventions.
 \begin{verbatim}
nparam = {-8.0339,1.07673,13.2425,1.96639,7.71759,7.57298,13.4593,3.13037, 
3.5639,3.83348,4.05641,4.22857,4.48498,4.71726,4.92507,5.83047};

vparam = {x0,y0,r1,r2,r3,r4,rcx,rcy,psi1,psi2,psi3,psi4,psi5,psi6,psi7,gamma};

supersolanima = Thread[Rule[vparam, nparam]]

r0 = {x0, y0} /. supersolanima;

xs = {7.03,6.95,6.77,6.4,5.91,5.43,4.93,4.67,4.38,4.04,3.76,3.76,3.76,3.76, 
3.76,3.76,3.76,3.8,4.07,4.53,5.07,5.05,5.89,6.41,6.92};

ys = {5.99,5.45,5.03,4.6,4.03,3.56,2.94,2.6,2.2,1.657,1.22,1.97,2.78,3.56, 
4.34,4.91,5.47,5.98,6.4,6.75,6.85,6.84,6.83,6.8,6.58};

DatT = Thread[{xs, ys}];
Dat = Take[DatT, {11, 25}];

L3 = (r4^2 - r1^2 - r2^2 - r3^2)/(2 r2 r3);
L2 = r1/r3;
L1 = r1/r2;

KA = Cos[psi] - L1 + L2 Cos[psi] + L3;
KB = -2 Sin[psi];
KC = L1 + (L2 - 1) Cos[psi] + L3;

theta[psi_] = 2 ArcTan[(-KB - Sqrt[KB^2 - 4 KA KC])/
(2 KA)] /.supersolanima;

Px[psi_] = (x0 + Cos[gamma] (r2 Cos[psi] + rcx Cos[theta[psi]] - 
rcy Sin[theta[psi]]) - Sin[gamma] (r2 Sin[psi] + rcx Sin[theta[psi]] + 
rcy Cos[theta[psi]])) /.supersolanima;

Py[psi_] = (y0 + Sin[gamma] (r2 Cos[psi] + rcx Cos[theta[psi]] - 
rcy Sin[theta[psi]]) + Cos[gamma] (r2 Sin[psi] + rcx Sin[theta[psi]] +
rcy Cos[theta[psi]])) /.supersolanima;

PrPl[psi_] = Thread[{Px[psi], Py[psi]}];

B[psi_] = r0 + {r2 Cos[psi + gamma], r2 Sin[psi + gamma]} /.supersolanima;

Cc[psi_] = B[psi] + {r3 Cos[theta[psi] + gamma], r3 Sin[theta[psi] +
gamma]} /.supersolanima;

CoorD = (r0 + r1 {Cos[gamma], Sin[gamma]}) /. supersolanima;

gDat = ListPlot[Dat];

linkb[psi_] := Graphics[{Thick, Line[{B[psi], r0}]}];
linkc[psi_] := Graphics[{Thick, Line[{Cc[psi], B[psi]}]}];
linkd[psi_] := Graphics[{Thick, Line[{Cc[psi], CoorD}]}];
linke[psi_] := Graphics[{Thick, Line[{PrPl[psi], B[psi]}]}];
linkf[psi_] := Graphics[{Thick, Line[{PrPl[psi], Cc[psi]}]}];

linka = Graphics[{Thickness[.01], EdgeForm[Thick],RGBColor[0.75, 0.75, 0.75],
Polygon[{r0, {r0[[1]], CoorD[[2]]}, CoorD}]},PlotRange -> {{-10, 10}, 
{-4.8, 8}}];

gr = ListPlot[{Table[PrPl[psi], {psi, 0, 2 Pi, Pi/50}], Dat},Joined -> 
{True, False}, PlotStyle ->{{PointSize[Medium],AbsoluteThickness[1.1]}},
PlotRange ->{{-10, 10}, {-5.2, 7.5}},  Frame -> True, Axes -> False, 
FrameLabel -> {"X", "Y"}, AspectRatio -> Automatic];

Animate[Show[{linka, gr, linkb[psi], linkc[psi], linkd[psi], linke[psi],
linkf[psi], gDat}, Axes -> True, AspectRatio -> Automatic], {psi, 0, 2 Pi}]
 \end{verbatim}